\documentclass{article}
\usepackage{spconf,amsmath,graphicx}
\usepackage{url}
\usepackage{multirow}
\usepackage{subfig}
\usepackage{epstopdf}
\title{Robustness of Voice Conversion Techniques Under Mismatched Conditions}

\name{Monisankha Pal$^{1}$, Dipjyoti Paul $^{1}$, Md Sahidullah$^{2}$, Goutam Saha$^{1}$}
\address{$^{1}$Department of E \& ECE, Indian Institute of Technology Kharagpur, Kharagpur, India\\
$^{2}$School of Computing, University of Eastern Finland, Joensuu, Finland\\
e-mail: \small monisankhapal@iitkgp.ac.in, dipjyotipaul@ece.iitkgp.ernet.in, sahid@cs.uef.fi, gsaha@ece.iitkgp.ernet.in}

\begin{document}
\maketitle

\begin{abstract}
Most of the existing studies on voice conversion (VC) are conducted in acoustically matched conditions between source and target signal. However, the robustness of VC methods in presence of mismatch remains unknown. In this paper, we report a comparative analysis of different VC techniques under mismatched conditions. The extensive experiments with five different VC techniques on CMU ARCTIC corpus suggest that performance of VC methods substantially degrades in noisy conditions. We have found that bilinear frequency warping with amplitude scaling (BLFWAS) outperforms other methods in most of the noisy conditions. We further explore the suitability of different speech enhancement techniques for robust conversion. The objective evaluation results indicate that spectral subtraction and log minimum mean square error (logMMSE) based speech enhancement techniques can be used to improve the performance in specific noisy conditions.
\end{abstract}
\vspace{-5pt}
\section{Introduction}
\vspace{-4pt}
\label{sec:intro}
Voice conversion (VC) is a methodology applied to a source speaker's speech signal to convert speaker identity. It creates the perception as if spoken by a specified target speaker while keeping the linguistic content unchanged. It has a wide variety of applications in text-to-speech (TTS) customization, designing of speaking aids, film dubbing, entertainment. There are large number of statistical approaches to VC like linear multivariate regression (LMR) \cite{valbret}, Gaussian mixture model (GMM) \cite{sty} etc. Over the time, the \emph{joint density GMM} (JDGMM) \cite{kain} based statistical parametric VC became the de facto standard and popular method. Later, other methods based on partial least squares (PLS) regression \cite{helander} and noisy channel model \cite{saito} have also been proposed. Furthermore, many non-linear spectral mapping techniques based on artificial neural network (ANN) \cite{desai}, dynamic kernel PLS regression \cite{helander1} and deep neural network (DNN) \cite{chen} have been developed.
\par
The statistical JDGMM based mapping produces converted speech with good conversion score, but suffers from over-smoothing and over-training effects \cite{toda}. This leads to degraded quality of speech. On the contrary, \emph{dynamic frequency warping} (DFW) based VC provides high quality converted speech, whereas the similarity to target speaker is low \cite{valbret}. In order to generate a trade-off solution, several hybrid methods such as \emph{weighted frequency warping} (WFW) \cite{erro}, DFW with amplitude scaling \cite{godoy}, \emph{bilinear frequency warping with amplitude scaling} (BLFWAS) \cite{erro2} have been investigated. The intrinsic problems of over-smoothing and over-training in probabilistic JDGMM based VC were overcome by adopting several techniques like Eigenvoice VC \cite{toda1}, \emph{mixture of factor analyzers} (MFA) \cite{wu1} and exemplar-based sparse representation \cite{wu}.
\vspace{-2pt}
\par
It is found that all these VC approaches provide reasonable performance with clean speech data. However, to the best of our knowledge, their effectiveness in \emph{noisy conditions} and comparative evaluation have not been studied yet. Meanwhile, some noise robust VC methods that use sparse non-negative matrix factorization (NMF) \cite{aihara2014noise}, affine NMF \cite{aihara2015small} and exemplar based \cite{takashima2012exemplar} approach, were proposed. But, those methods consider background noise in input source speech during test. In this paper, we consider the practical scenario where the source and target for voice conversion are not available from same environmental condition. In real world deployment of VC technology, it is always not possible to record target speaker's speech in sound proof booth with the presence of very low environmental noise. Whereas, the source speaker's voice can be collected in a controlled environment. The contribution of this work is a detailed analysis of five different popular VC methods in presence of white, babble and volvo noises, with three different signal-to-noise ratio (SNR) levels in target speech training data. We further examine the effectiveness of standard \emph{speech enhancement} methods as a post-processing module to reduce the mismatch.

\vspace{-10pt}
\section{MOTIVATION OF THE WORK}
\vspace{-10pt}
\label{sec:format}
One of the important practical concern in voice conversion is that during training, target speaker's speech samples (parallel or non-parallel) may not be available with the same acoustic condition as source speaker's voice. Moreover, sometimes source speaker who is trying to mimic the target speaker by using voice conversion algorithms, may get target data from different sources like audio clippings, TV, internet resources, etc. Naturally, the effects of mismatch due to the variation in the channel, handset and background noises exist. Most of the existing studies in this field have considered source and target speaker's voice from similar recording environment where the mismatch is not taken into account. As, for real world application of VC, this is very important to know the robustness of existing methods, we consider evaluating their performance in presence of noise.

\begin{figure}[t]
 \centering
  \includegraphics[height=2.5cm,width=0.52\textwidth]{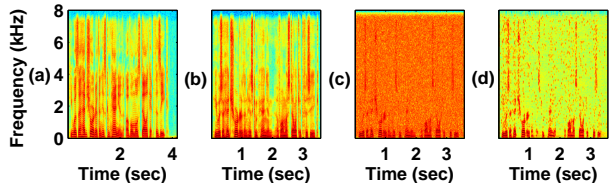}\\
  \vspace{-0.35cm}
  \caption{Spectrograms of the same utterance for (a) source, (b) target, (c) noisy target (0 dB, white) and (d) enhanced noisy target by spectral subtraction speech enhancement method.}\label{fig1}
  \vspace{-0.5cm}
\end{figure}

\par
We illustrate the spectrograms of clean source and target speech along with noisy target speech using white noise in Fig. \ref{fig1}. The differences in spectrogram are clearly observed between source and target and it should affect the mapping function used in VC algorithm. The spectrogram of enhanced version of the noisy speech by using spectral subtraction is also illustrated in Fig. \ref{fig1}. It is also not confirmed whether applying standard speech enhancement techniques will improve the VC performance or not, as they introduce some sort of speech distortion \cite{hanilci2016spoofing}. Therefore, we further explore the impact of speech enhancement methods on voice conversion. Finally, our study is to provide a clear insight into the robustness of different VC methods against noisy conditions. It also targets to give a solution by employing speech enhancement methods which can act as a test bench to demand for a new noise robust VC method in future.
\vspace{-5pt}
\section{Voice conversion with integrated speech enhancement}
\label{sec:pagestyle}
\begin{figure}[t]
 \centering
  \includegraphics[height=4cm,width=0.5\textwidth]{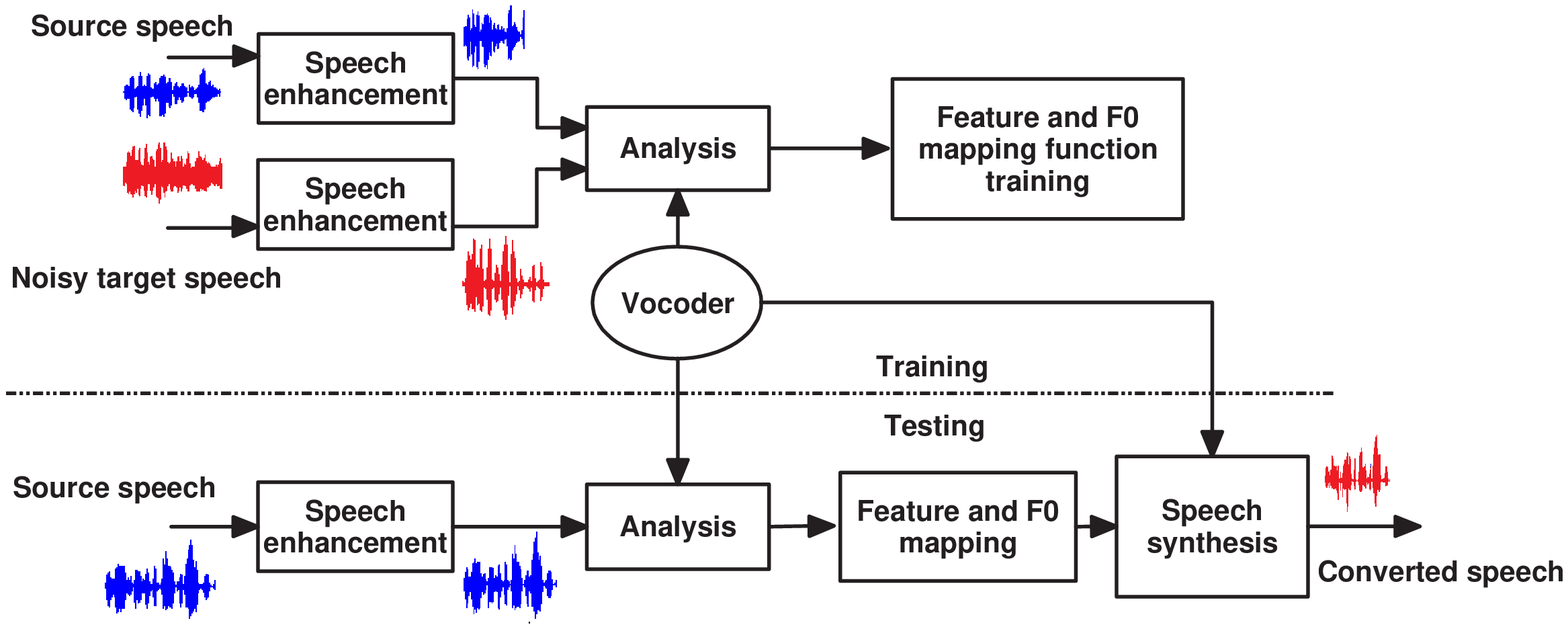}\\
  \vspace{-10pt}
  \caption{Block diagram of the evaluation framework.}\label{fig2}
  \vspace{-10pt}
\end{figure}

\vspace{-10pt}
Fig. \ref{fig2} shows the framework for robust voice conversion. In this framework, we study the effect of speech enhancement techniques. Three standard speech enhancement algorithms namely \emph{spectral subtraction} \cite{boll1979suppression}, \emph{iterative Wiener filtering} \cite{lim1979enhancement} and \emph{logarithmic minimum mean square error} (logMMSE) \cite{ephraim1985speech} are incorporated. Among them, the first one is spectrum based and the rest are statistical approaches for speech enhancement. Note that all these methods are applied to all speech signals including training and testing. During training, speech signals of source and target speakers are analyzed by using harmonic plus stochastic model (HSM) \cite{erro} or STRAIGHT \cite{kawahara1999restructuring} vocoder. Then, they are parameterized to either LSF or MFCC feature. After that, source and target features are aligned to develop a mapping function in VC training, which is used in the conversion stage to map source speaker's feature space to that of target speaker. Finally, the converted source features are synthesized using the mentioned vocoders.

\vspace{-12pt}
\section{VOICE CONVERSION Methods}
\label{sec:majhead}
\vspace{-5pt}
\subsection{Joint Density GMM (JDGMM)}
\vspace{-5pt}
In this method, after frame alignment, the joint density of the concatenated source and target feature vectors is modelled by a GMM \cite{kain,erro,pal2015robustness}. During conversion phase, given the source feature vector, the target vector is predicted using Gaussian regression with minimum mean square error (MMSE) criteria \cite{erro}.
\vspace{-12pt}
\subsection{Weighted Frequency Warping (WFW)}
\vspace{-5pt}
For each acoustic class, while training, the corresponding piece-wise linear frequency warping function estimation is described in \cite{erro}. In the conversion stage, a warping function for input source frame is calculated. After that, the input source frame spectral envelope is warped in frequency. The final converted spectrum is obtained as the multiplication between an energy correction filter response and the warped source spectral envelope. The energy correction filter is used to correct the overall energy distribution of spectrum and spectral tilt.
\vspace{-12pt}
\subsection{Dynamic Frequency Warping (DFW)}
\vspace{-5pt}
In this method, spectral mapping follows the frequency warping based approach with energy correction filter disabled. Therefore, this method is similar to the previous one except that no energy correction is applied \cite{valbret,erro}.
\vspace{-12pt}
\subsection{Mixture of Factor Analyzer (MFA)}
\vspace{-5pt}
In this method, speaker independent phonetic vectors and factor loadings are estimated from non-parallel prior data in off-line \cite{wu1}. Although, this method needs some parallel training data, it is a reliable model due to off-line training and \emph{tied} nature of the covariance matrix. During VC training, speaker dependent identity vectors and \emph{tied} covariance matrix are determined. In the conversion phase, the target feature vector is calculated using the conversion function given in \cite{wu1}.
\vspace{-12pt}
\subsection{Bilinear Frequency Warping with Amplitude Scaling (BLFWAS)}
\vspace{-5pt}
This VC method combines bilinear frequency warping (BLFW) and amplitude scaling (AS) \cite{erro2}. BLFW is based on a single parameter to reduce the number of parameters to be estimated using a small amount of target training data. Therefore, it is robust against target data scarcity and less prone to \emph{over-fitting}. In order to reduce the difference between target and warped source spectra, amplitude scaling was adopted.
\vspace{-5pt}
\section{Database and performance evaluation}
\vspace{-5pt}
\label{sec:typestyle}
\subsection{Speech corpus}
\vspace{-5pt}
For the performance evaluation of different VC techniques, experiments are conducted on CMU ARCTIC \cite{kominek2004cmu} corpus. The statistical details of the data setup to perform our VC experiments are described in Table \ref{table1}. We have digitally added white, babble and volvo noises from NOISEX-92 \cite{varga1993assessment} corpus with 0 dB, 10 dB and 20 dB SNR levels to each target training speech sample using filtering and noise adding tool (FANT) \footnote{\scriptsize \url{http://dnt.kr.hsnr.de/}}. This open source tool follows ITU recommendations for noise adding and filtering. The speech and noise sampling frequency for all the cases are 16 kHz.
\par
For the first three VC methods, the speech files are analyzed using HSM vocoder at a rate of 128 samples per frame. LSF feature of order 14 is used for spectral parameterizations. GMM of model order 8 with covariance matrix type \emph{full} is constructed for all these methods. For the last two VC methods, STRAIGHT analysis/synthesis framework is adopted. It decomposes the speech signal into F0 and spectral envelope at 5 ms steps. MFCCs of order 24 are employed to represent the spectral envelope. In MFA VC method, GMM of model order 128 is obtained in off-line using non-parallel prior data from TIMIT corpus \footnote{\scriptsize \url{https://catalog.ldc.upenn.edu/LDC93S1}}. The covariance matrix type considered for prior GMM and VC training are \emph{diag} and \emph{tied}, respectively. Moreover, a \emph{full} covariance GMM of order 16 is used for the last VC. In all the VC systems, F0 is linearly converted using mean-variance equalization of log F0 of source and target speaker's training data.

\vspace{-5pt}
\subsection{Performance metrics}
\vspace{-5pt}

Mel cepstral distortion (MCD) \cite{helander1,wu1} and PESQ \cite{hu2008evaluation} are incorporated as performance metrics for the objective evaluation of voice converted wavefiles. MCD is used to measure the spectral mapping performance, while PESQ is meaningful for objective quality assessment \cite{huang2011prediction}. The MCD between original target and converted mel cepstra is determined as
\begin{equation}
\scriptsize
MCD_{t}(dB) = \frac{10}{ln 10} \sqrt{2 \displaystyle\sum_{i=1}^{24} \left(mc_{t}^{i}- \hat{mc}_{t}^{i}\right)^2}
\end{equation}
where, $mc_{t}^{i}$ and $\hat{mc}_{t}^{i}$ denote the $i$-th dimension MFCCs of target and converted features at frame $t$. The lower the MCD value is, the better the conversion is. On the other hand, PESQ metric is recommended by ITU-T for speech quality assessment. It is computed using a linear combination of the average normal disturbance value $d_{sym}$ and the average asymmetrical disturbance value $d_{asym}$ between the reference and converted loudness spectra as: $ \scriptsize PESQ=4.5-0.1 \times d_{sym}-0.0309 \times d_{asym}$. The detail of disturbance values calculation are given in \cite{hu2008evaluation}. The higher the PESQ value is, the better the speech quality is.
\vspace{-2pt}
\begin{table}[t]
\renewcommand{\arraystretch}{1.3}
\setlength{\tabcolsep}{1.2pt}
\centering
\scriptsize
\caption{\small CMU ARCTIC corpus description for VC experiments.}
\vspace{-0.2cm}
\label{table1}
\begin{tabular}{|c|c|c|c|c|c|}
\hline
\multirow{2}{*}{Subset} & \multicolumn{2}{c|}{No of speakers}                       & \multirow{2}{*}{No of utt.} & \multirow{2}{*}{\begin{tabular}[c]{@{}c@{}}No of conversion\\directions\end{tabular}} & \multirow{2}{*}{\begin{tabular}[c]{@{}c@{}}Total no of \\ test utt.\end{tabular}} \\ \cline{2-3}
                        & Male                          & Female                        &                                       &                                                                                             &                                                                                             \\ \hline
Train                & 2 & 2 & 30 (1--30)                       & 12 (all possible                                                                          & \multirow{2}{*}{240 (20 $\times$ 12)}                                                                \\
\cline{1-1} \cline{4-4}
Test                 &    (BDL, RMS)                           &      (CLB, SLT)                         & 20 (51--70)                      &   speaker pairs)                                                                                       &                                                                                             \\ \hline
\end{tabular}
\vspace{-0.45cm}
\end{table}

\begin{table*}[t]
\vspace{-2.3pt}
\renewcommand{\arraystretch}{1.5}
\setlength{\tabcolsep}{2pt}
\centering
\scriptsize
\caption{\small Average MCD and PESQ scores in clean as well as noisy target conditions for five different VC methods.}
\vspace{-0.25cm}
\label{table2}
\begin{tabular}{|c|c|ccc|ccc|ccc||c|ccc|ccc|ccc|}
\hline
\multirow{3}{*}{VC type} &  \multicolumn{10}{c||}{Average MCD (dB) } &  \multicolumn{10}{c|}{Average PESQ }                                                       \\ \cline{2-11} \cline{11-21}
                         & \multirow{2}{*}{Clean}&\multicolumn{3}{c|}{White}  & \multicolumn{3}{c|}{Babble} & \multicolumn{3}{c||}{Volvo} & \multirow{2}{*}{Clean}&\multicolumn{3}{c|}{White}  & \multicolumn{3}{c|}{Babble} & \multicolumn{3}{c|}{Volvo}\\ \cline{3-11} \cline{13-21}
                         & & 0 dB  & 10 dB & 20 dB & 0 dB  & 10 dB & 20 dB & 0 dB & 10 dB & 20 dB & & 0 dB  & 10 dB & 20 dB & 0 dB  & 10 dB & 20 dB & 0 dB & 10 dB & 20 dB\\ \hline
DFW                      & 7.62&\textbf{8.53}  & 8.19  & 7.93  & 8.34  & 7.96  & 7.78  & 7.94 & 7.81  & 7.73  & 0.89&\textbf{0.73} & \textbf{0.82}  & 0.86  & \textbf{0.69}  & 0.87  & 0.89  & 0.76 & 0.81  & 0.90\\
JDGMM                    & 7.36&9.10  & 8.46  & 7.89  & 8.47  & 7.92  & 7.59  & 7.84 & 7.59  & 7.44 & 0.91&0.61 & 0.77  & 0.85  & 0.60  & 0.79  & 0.86  & 0.63 & 0.75  & 0.90 \\
WFW                      &7.60 &9.27  & 8.59  & 8.04  & 8.63  & 8.11  & 7.80  & 7.96 & 7.74  & 7.66  & 0.96&0.60 & 0.76  & 0.89  & 0.65  & 0.85  & 0.90  & 0.75 & 0.85  & 0.94\\
MFA                      & \textbf{6.84}&11.70 & \textbf{8.17}  & \textbf{7.35}  & 8.24  & 7.99  & 7.63  & 9.44 & 8.36  & 7.56 &\textbf{1.03} &0.27 & 0.70  & \textbf{0.96}  & 0.61  & \textbf{0.92}  & \textbf{1.02}  & 0.68 & 0.91  & 1.01 \\
BLFWAS                    & 7.07&8.81  & 8.42  & 7.73  & \textbf{8.17}  & \textbf{7.57}  & \textbf{7.17}  & \textbf{7.57} & \textbf{7.16}  & \textbf{7.07} &0.99 &0.48 & 0.47  & 0.80  & 0.60 & 0.88  & 1.00  & \textbf{0.82} & \textbf{0.97}  & \textbf{1.03} \\ \hline
\end{tabular}
\vspace{-0.25cm}
\end{table*}

\begin{table*}[]
\vspace{-1pt}
\renewcommand{\arraystretch}{1.5}
\setlength{\tabcolsep}{2pt}
\centering
\scriptsize
\caption{\small Average MCD and PESQ scores with spectral subtraction, iterative Wiener filtering and logMMSE speech enhancement method employed in clean as well as noisy target conditions for five different VC methods.}
\vspace{-0.23cm}
\label{table3}

\begin{tabular}{|c|c|c|ccc|ccc|ccc||c|ccc|ccc|ccc|}

\hline
\multirow{3}{*}{Speech}&\multirow{3}{*}{ VC type} &  \multicolumn{10}{c||}{Average MCD (dB) } &  \multicolumn{10}{c|}{Average PESQ }                                                       \\ \cline{3-12} \cline{12-22}
                    \multirow{2}{*}{enhancement}   &  & \multirow{2}{*}{Clean}&\multicolumn{3}{c|}{White}  & \multicolumn{3}{c|}{Babble} & \multicolumn{3}{c||}{Volvo} & \multirow{2}{*}{Clean}&\multicolumn{3}{c|}{White}  & \multicolumn{3}{c|}{Babble} & \multicolumn{3}{c|}{Volvo}\\ \cline{4-12} \cline{14-22}
                         & & &0 dB  & 10 dB & 20 dB & 0 dB  & 10 dB & 20 dB & 0 dB & 10 dB & 20 dB & & 0 dB  & 10 dB & 20 dB & 0 dB  & 10 dB & 20 dB & 0 dB & 10 dB & 20 dB\\ \hline
Spectral & DFW                      & 7.71&8.63  & 7.99  & 7.90  & 9.05  & 8.00  & 7.77  & 8.08 & 7.80  & 7.71  & 0.92&0.59 & 0.84  & 0.86  & 0.47  & 0.84  & 0.88  & 0.73 & 0.89  & 0.90\\
subtraction&JDGMM                    & 7.39&8.98  & 7.84  & 7.62  & 9.30  & 7.79  & 7.51  & 7.99 & 7.54  & 7.43 & 0.91&0.55 & 0.88  & 0.91  & 0.40  & 0.83  & 0.91  & 0.76 & 0.91  & 0.92 \\
&WFW                      &7.65 &9.13  & 8.04  & 7.89  & 9.54  & 8.01  & 7.72  & 8.15 & 7.69  & 7.65  & 0.93&0.52 & 0.89  & 0.92  & 0.41  & 0.86  & 0.92  & 0.75 & 0.92  & 0.96\\
&MFA                      & \textbf{7.10}& \textbf{8.40} & \textbf{7.69} & \textbf{7.34}  & 8.88  & 8.29  & 7.77  & 8.40 & 7.98  & 7.36 &\textbf{1.04} &\textbf{0.91} & \textbf{1.00}  & \textbf{1.01}  & \textbf{0.71}  & \textbf{0.95}  & 1.03  & \textbf{0.83} & 0.95  & \textbf{1.06} \\
&BLFWAS                    & 7.26&8.53& 7.89  & 7.46  & \textbf{8.56}  & \textbf{7.62}  & \textbf{7.33}  & \textbf{7.86} & \textbf{7.48}  & \textbf{7.26} &\textbf{1.04}&0.58 & 0.90  & 0.96  & 0.67  & 0.93  & \textbf{1.06}  & 0.82 & \textbf{0.96}  & 1.03 \\ \hline
Iterative&DFW                      & 8.59&9.21  & 9.08  & 8.77  & \textbf{9.84}  & 8.78  & 8.68  & 8.86 & 8.66  & 8.60  & 0.81&0.76 & 0.78  & 0.84  & 0.54  & 0.77  & 0.83  & 0.71 & 0.84  & 0.84\\
Wiener&JDGMM                    & \textbf{8.22}&\textbf{9.11}  & \textbf{8.79}  & \textbf{8.48}  & 9.97  & \textbf{8.59}  & \textbf{8.32}  & \textbf{8.79} & \textbf{8.40}  & \textbf{8.30} & 0.84&0.81 & 0.89  & 0.89  & 0.55  & 0.81  & 0.85  & \textbf{0.85} & \textbf{0.92}  & 0.90 \\
filtering&WFW                      &8.54 &9.22  & 9.00  & 8.70  & 10.22  & 8.86  & 8.64  & 8.97 & 8.69  & 8.61  & 0.89&0.82 & 0.88  & 0.90  & 0.54  & 0.83  & 0.86  & 0.74 & 0.83  & 0.88\\
&MFA                      & 9.46&11.60 & 11.07 & 10.49  & 10.99  & 9.99  & 9.73  & 10.99 & 9.68  & 9.49 &\textbf{0.95} &\textbf{1.02} & \textbf{1.02}  & \textbf{0.96}  & \textbf{0.70}  & \textbf{0.88}  & \textbf{0.91}  & 0.63 & \textbf{0.92}  & \textbf{0.95} \\
&BLFWAS                    & 8.47&9.69& 9.32  & 8.81  & 10.05  & 8.81  & 8.53  & 9.25 & 8.68  & 8.48 &0.88&0.93 & 0.92  & 0.93  & 0.69  & 0.87  & 0.87  &0.68  & 0.84  & 0.88 \\ \hline
Log&DFW                      & 7.79&8.29  & 7.97  & 7.89  & \textbf{8.43}  & 7.93  & 7.79  & 7.91 & 7.79  & 7.79  & 0.87&0.81 & 0.86  & 0.89  & \textbf{0.66}  & 0.83  & 0.90  & 0.79 & 0.86  & 0.88\\
MMSE&JDGMM                    & \textbf{7.49}&\textbf{8.23}  & \textbf{7.86}  & \textbf{7.59}  & 8.54  & \textbf{7.76}  & \textbf{7.50}  & \textbf{7.77} & \textbf{7.56}  & \textbf{7.49} & 0.90&0.80 & 0.89  & 0.89  & 0.63  & 0.82  & 0.90  & \textbf{0.83} & \textbf{0.92}  & 0.91 \\
&WFW                      &7.73 &8.47  & 8.03  & 7.81  & 8.81  & 7.92  & 7.71  & 7.93 & 7.74  & 7.75  & \textbf{0.91}&\textbf{0.83} & \textbf{0.90}  & \textbf{0.95}  & 0.61  & \textbf{0.87}  & \textbf{0.92}  & 0.78 & 0.90  & \textbf{0.93}\\
&MFA                      & 8.82&8.81 & 8.80 & 8.78  & 8.88  & 8.78  & 8.78  & 8.88 & 8.80  & 8.80 &0.83 &0.79 & 0.82  & 0.82  & 0.62  & 0.76  & 0.81  & 0.66 & 0.75  & 0.79 \\
&BLFWAS                    & 8.76&8.81& 8.80  & 8.78  & 8.88  & 8.78  & 8.77  & 8.87 & 8.80  & 8.81 &0.85&0.79 & 0.82  & 0.82  & 0.62  & 0.76  & 0.81 & 0.66 & 0.75  & 0.80 \\ \hline
\end{tabular}
\vspace{-0.45cm}
\end{table*}

\vspace{-5pt}
\section{RESULTS AND ANALYSIS}
\label{ssec:subhead}
\vspace{-5pt}
Experimental results on the performance of different VC methods under mismatched condition by using noisy target speech are presented in Table \ref{table2}. Average MCD and PESQ scores for objective similarity and quality are reported. As shown in this table, in original (clean) case, MFA VC is superior to other methods. Whereas, in noisy conditions, the performance of all the methods deteriorates. This could be due to difficulty in estimating F0s from noisy speech. It is also clear that whenever the target training data is not clean, all the mapping functions fail to learn the correspondence between source and target acoustic features. For white noise with high SNR levels (10 dB, 20 dB), MFA yields lowest MCD score. DFW produces lowest MCD in case of white noise with 0 dB SNR level. The BLFWAS VC method performs well for both babble and volvo noises. The probable reason could be the fact that both the MFA and BLFWAS methods are more robust against target data scarcity condition. They are less affected by over-fitting. Due to over-fitting, the predictive performance is poorer for the other three methods and it is further degraded in presence of noise. On the other hand, we get lower MCD scores in volvo noise than babble and white noises, as it affects only the lower frequency region of the target spectrum. DFW provides superior performance in terms of average PESQ value than  other methods in 0 dB SNR values. JDGMM yields low PESQ value for almost all the cases due to its over-smoothing mapping procedure and in presence of noise, naturalness further reduces. MFA yields better PESQ values for high SNR levels of noise. BLFWAS outperforms other methods as a whole, both in MCD and PESQ values. Finally, it is also worthwhile to mention that MFCC feature is more robust in additive noise as compared to LSF feature, which is the case for last two VC techniques of the table.
\par
Voice conversion performances with integrated speech enhancement technique are shown in Table \ref{table3}. For spectral subtraction it shows that for all the VC methods, we obtain similar or slightly degraded performance in clean case as compared to Table \ref{table2}. For the case of MCD score, it is observed from the table that in case of white noise with all the SNR levels, the performance improves for all VC methods. However, no significant performance gain is noticed for babble and volvo noises, except MFA in 0 dB white and volvo noise. The table also shows similar kind of pattern for PESQ score like MCD, except the MFA method. Here, the performance improvement in PESQ value is more prominent than other techniques.
\par
Table \ref{table3} also presents the average MCD and PESQ values when iterative Wiener filtering is incorporated as speech enhancement method. It can be seen from the table that the performance degrades for clean and all noisy cases and also for all VC methods. This is probably because it introduces some processing artifacts and that is why input speech is less successfully converted. However, in case of PESQ score, almost all the methods yield performance improvement in white noise and no significant improvements for the other two noises except JDGMM in volvo noise. As a whole, JDGMM provides superior performance as compared to other methods.
\par
Finally, in Table \ref{table3}, the logMMSE speech enhancement is applied. Here, no significant performance gain in MCD score is observed. The only exception is JDGMM in white noise. The PESQ value increases for all the systems in white noise. For DFW, WFW and JDGMM methods, PESQ score slightly increases for almost all the noisy cases. JDGMM outperforms other VC methods when we employ logMMSE based speech enhancement.
\vspace{-15pt}
\section{CONCLUSION}
\vspace{-10pt}
This study presents a detailed analysis of the robustness of existing VC methods against mismatched condition with noisy target data. It reveals that in most of the cases, BLFWAS provides superior performance than other VC methods. However, in white noise, MFA outperforms others. We also explored the effectiveness of speech enhancement methods on all the speech samples in clean as well as noisy cases. We have observed that spectral subtraction improves performance in case of white noise while iterative Wiener filtering degrades the performance. Furthermore, log MMSE provides no performance gain in MCD value. However, it gives better PESQ in white noise. As a whole, spectral subtraction works well for BLFWAS and MFA, while iterative Wiener filtering and logMMSE for JDGMM. As future work, we plan to extend this analysis for recently developed DNN-based voice conversion techniques. The results from this work could be useful for the development of robust voice conversion algorithm for real-world application.
\vspace{-10pt}

\section{\scriptsize Acknowledgement}
\vspace{-5pt}
\scriptsize
This work is funded by Ministry of HRD, Govt of India. The contribution of the third author is sponsored by the OCTAVE Project (\#647850), funded by the Research European Agency (REA) of the European Commission, in its framework programme Horizon 2020. The views expressed in this paper are those of the authors and do not engage any official position on the European Commission.
\vfill\pagebreak

\label{sec:refs}

\fontsize{8.3pt}{7}\selectfont

\bibliographystyle{IEEEbib}
\bibliography{ref1}

\end{document}